\documentclass[11pt]{article}
\usepackage{amsmath, amssymb, amsthm, amsfonts, amssymb, graphicx, epstopdf} 
\usepackage{amsmath}
\usepackage{amsthm}
\usepackage{cite}
\usepackage{subfig}
\usepackage{array}
\usepackage{multirow}
\usepackage{longtable}
\usepackage{multicol}
\usepackage{cite}
\usepackage{float}
\usepackage{makecell}
\usepackage{wasysym,amsbsy}
\usepackage[table]{xcolor}
\usepackage{geometry}
\usepackage{xspace}
\usepackage[mathscr]{euscript}
\usepackage{mathtools}
\usepackage{hyperref}
\usepackage[utf8]{inputenc}
\usepackage[english]{babel}
\usepackage{mathtools}
\usepackage{lscape}
\usepackage{hyperref}
\usepackage[version=3]{mhchem}

\newcolumntype{P}[1]{>{\centering\arraybackslash}p{#1}}

\date{}

\begin{document}
	\date{}
	\title{\textbf{PyTIE: A Python Program for the Evaluation of Degree-Based Topological Descriptors and Molecular Entropy}}
	\author{Sahaya Vijay Jeyaraj$^{a}$, Roy S$^{a}$\footnote{corresponding author: roy.santiago@vit.ac.in}, Govardhan S$^{b}$, Tony Augustine$^{c}$, Jyothish K$^{a}$}
	\maketitle
	\vspace{-0.5cm}
	{\centering
		$^{a}$ Department of Mathematics, School of Advanced Sciences, Vellore Institute of Technology, Vellore-632014, India \\
		$^{b}$ Department of Mathematics, Rajalakshmi institute of technology, Chennai-600124, India \\
		$^{c}$ Department of Mathematics, Nirmala College (Autonomous), Muvattupuzha,\\ Ernakulam-686661, Kerala, India \\
	}
	\begin{abstract}
		We have developed PyTIE (Python Topological Indices Expressions) which is defined as the collections of Python packages such as $\text{PyTIE\_D}$, $\text{PyTIE\_DS}$, $\text{PyTIE\_SMS\_DE}$, and $\text{PyTIE\_SMS\_DSE}$, which are open-source software packages and cross-platform Python package designed to expedite the retrieval of results for mathematics, chemistry and chemical engineering researchers within constant time. This open-source tool extends its utility to chemistry and chemical engineering researchers with limited mathematical proficiency. PyTIE facilitates the loading of molecular graphs, specifying parameters such as minimum degree, maximum degree, and the number of vertex pairs (edge partitions). The edge partitions of a molecular graph based on degree sum also plays a crucial role in predicting heat of formation and enthalpy of formation along with DFT techniques. It systematically computes expressions and numerical values for various topological indices, including degree-based and neighborhood degree-based indices, as well as Shannon’s entropy, providing visual representations of the results. Emphasizing topological indices for Quantitative Structure-Activity Relationship and Quantitative Structure-Property Relationship analyses, PyTIE proves particularly relevant in these studies. Serving as a Python package, it seamlessly integrates with libraries such as NumPy, math and SymPy offering extensive options for data analysis. The efficiency of PyTIE is demonstrated through illustrative examples in various contexts.
	\end{abstract}
	\section{Introduction}	
	In chemistry, molecular descriptors numerical functions generated from molecular structures are extremely important, especially in QSAR and QSPR investigations \cite{pytie18,pytie19,pytie20}. They establish a link between specific molecular descriptors and the biological or chemical properties of molecules, making it possible to anticipate qualities based solely on structure and obviating the necessity for synthesis. One important class of molecular descriptors that originates from molecular graphs (usually with suppressed hydrogen) is topological indices, which omit information such as angles and bond lengths \cite{pytie21,pytie22}. They encode information about atom adjacencies and branching inside a molecule, acting as graph invariants. Examples of well-known topological indices include Wiener index, Zagreb index and Randi\'c index \cite{pytie23,pytie24,pytie25}. 
	
	Topological indices provide information on physical and chemical properties by studying the connectivity patterns of atoms within a molecule. This data is useful for forecasting molecular behavior, measuring reactivity, and optimizing synthesis pathways \cite{pytie47,pytie48}. 
	In materials research, topological indices serve as fundamental tools in guiding and optimizing the design and development of novel materials by forecasting molecular interactions under diverse circumstances. By forecasting biological activity, they aid in the identification of viable drug candidates in the pharmaceutical industry \cite{pytie49,pytie50,pytie51,pytie53,pytie61,pytie62}. Applications of Shannon's entropy can be found in many fields, including statistics, engineering, quantum computing, physics, and information theory \cite{pytie54,pytie55}. It is demonstrated in \cite{pytie56} that these entropies are helpful metrics for determining the thermodynamic stability of different chemical compounds. Artificial intelligence (AI) has advanced significantly in recent years because of deep learning and big data, which can now manage more extensive datasets. In 20222, K.J. Gowtham introduced the link among degree-based indices with degree-based entropies \cite{pytie57}.
	
	PyTIE is defined as a collections of Python software packages, are $\text{PyTIE\_D}$, $\text{PyTIE\_DS}$, \\ $\text{PyTIE\_SMS\_DE}$, and $\text{PyTIE\_SMS\_DSE}$ are vailable as open source, offering optimized transforms and a modular library of optimization tools for solving linear chemical structure issues. Here $\text{PyTIE\_D}$, $\text{PyTIE\_DS}$, $\text{PyTIE\_SMS\_DE}$, and $\text{PyTIE\_SMS\_DSE}$ stands follow Degree-based Python Topological Indices Expressions, Degree Sum-based Python Topological Indices Expressions,  Degree and Entropy based Python Topological Indices Expressions for a Single Molecular Structure, Degree Sum and Entropy based Python Topological Indices Expressions for a Single Molecular Structure. Originally tailored for QSPR and QSAR applications, the software's adaptability and the broad applicability of its mathematical techniques, emphasized in this study, indicate its efficiency and consistent time results in computational tasks \cite{pytie19,pytie20,pytie17,pytie59}.
	
	Existing QSAR software, including Dragon \cite{pytie2,pytie3}, Molgen-QSPR \cite{pytie4,pytie5}, GenerateMD \cite{pytie6}, PowerMV \cite{pytie7}, Molconn-Z \cite{pytie8}, CODESSA \cite{pytie9}, Chemical Descriptors Library \cite{pytie10}, AZOrange \cite{pytie11}, PaDEL-Descriptor \cite{pytie12,pytie13}, Chemistry Development Kit \cite{pytie14} and MathChem \cite{pytie17}, is primarily focused on calculating a limited set of well-known molecular descriptors. Many topological indices relevant to mathematical chemists are often overlooked or omitted. Dragon \cite{pytie2} boasts an extensive list of 4885 molecular descriptors, with over a thousand qualifying as topological indices, but it still lacks certain indices like Laplacian energy. 
	
	As a result of these considerations, PyTIE has been developed as an open-source Python package, serving as the sole software capable of providing expressions and numerical values for topological indices. Although Python \cite{pytie15} serves as a programming language, it adheres to a minimalist philosophy with a strong focus on code readability, evident in examples of PyTIE usage throughout the paper. Python’s qualities, including a brief learning curve, contribute to its widespread acceptance in the scientific community. Furthermore, PyTIE seamlessly integrates with numerous scientific software tools implemented in Python \cite{pytie15,pytie16}.
	\section{Preliminaries}
	Let $G$ represent the molecular graph of the chemical structure which are simple and connected. The symbols $E$ and $V$, which stand for the sets of edges and vertices of $G$, respectively, indicate the set of atoms and chemical bonds in the chemical system. The vertex's degree, represented by $d(v)$, is equal to the atomic valence. 
	In a graph $G$, the degree of the vertex with the highest number of incident edges is referred to as the maximum degree of $G$, denoted by $\Delta(G)$. On the other hand, the degree of the vertex with the lowest number of incident edges is termed the minimum degree of $G$, represented by $\delta(G)$.
	
	Furthermore,
	the collection of vertices adjacent to a vertex $v$ is denoted by $N(v)$, and $\displaystyle w(v)=\sum_{r \in N(v)} deg(r)$.
	Let $E$ be the edge connecting two nearby vertices, $u$ and $v$, with degrees $a$ and $b$, respectively. $E_{a,b}$ represents edge partition, which may be expressed as $E_{a,b} = \{deg(u), deg(v)\}$. 
	Let $E$ be the edge connecting two nearby vertices, $u$ and $v$, with degrees $a$ and $b$, respectively. $E_{a,b}$ represents edge partition, which may be expressed as $E_{a,b} = \{d(u), d(v)\}$. The purpose of this characterization is to examine the structural characteristics and connectivity of the molecular graph. 
	Shannon introduced the concept of entropy within the framework of information theory through probabilistic measures. This notion of entropy has subsequently been applied in graph theory to describe the probability distributions over the vertex and edge sets. In the context of chemical graph theory, such applications are essential for evaluating the system's free energy and assessing the stability of molecular orbitals.
	The average degree of surprise or information regarding potential outcomes of a discrete random variable is described by Shannon's entropy measure, which is utilized in information and coding theory. The following is the Shannon-defined entropy formula, 
	\begin{equation*}
		H(X) = -\sum_{i=1}^{T} P_i \log_2(P_i)
	\end{equation*}
	
	where $X$ denotes a collection of discrete random variables and $P_i$ represents their corresponding probability metrics.
	
	\begin{equation*}
		T = \sum_{j=1}^I |x_j|
	\end{equation*}
	
	where $I$ is the information's cardinality of $x_j$ and $T$ is the whole length of the total information.
	\section{Result and Discussion}
	\subsection{PyTIE Features}
	To address these issues and provide assistance for researchers in mathematical chemistry, we have developed PyTIE, which has the following features:
	\begin{enumerate}
		\item PyTIE’s capability to load molecular and ordinary graphs from both chemical and graph theoretical sources.
		\item PyTIE's efficiently computes topological indices of complex molecular structures with constant time complexity.
		\item PyTIE's provision of highly simplified expressions surpasses manual computations in terms of efficiency and accuracy.
		\item This capability not only conserves computational time but also diminishes the reliance on computational tools. 
		\item The accuracy and reliability of manual computational efforts can be verified through the utilization of PyTIE's computational software.
		\item Flexibility beyond solving predefined problem types.
		\item Ease of extension for anyone to add definitions of new topological indices to multiple parameters.
		\item In the event of erroneous input being provided to PyTIE modules, the system outputs the message "Invalid input detected."
		\item Class objects have been constructed within each PyTIE modules, as detailed below:
		\begin{table}[H]
			\centering
			\resizebox{2.5in}{.4in}{
				\begin{tabular}{|l|c|}
					\hline 
					{\bf Modules} & {\bf Class} \\
					\hline	
					PyTIE\_D & topological\_indices\_Degree\\ 
					\hline
					PyTIE\_DS & topological\_indices\_DegreeSum\\ 
					\hline
					PyTIE\_SMS\_DE & sms\_topological\_indices\_Degree\\
					\hline
					PyTIE\_SMS\_DSE & sms\_topological\_indices\_DegreeSum\\
					\hline
				\end{tabular}
			}
		\end{table}
		\item Functions have been constructed within each PyTIE modules, as detailed below:
		\begin{table}[H]
			\centering
			\caption{ Degree-based Topological indices, its formulas and its functions}\label{TIformula1}
			\resizebox{5.3in}{3.3in}{
				\begin{tabular}{|l|l|l|l|}
					\hline 
					S.No &{\bf Topological Indices}\cite{pytie33,pytie34,pytie35,pytie36,pytie37,pytie38,pytie39,pytie40,pytie41} & {\bf Formula} & {\bf Function}  \\
					\hline	
					1 & First Zagreb  ($M_{1}$)& $ \displaystyle M_1 = \sum_{uv \in E(G)} (deg_u + deg_v)$ & first\_Zagreb()\\
					2 & Second Zagreb  ($M_{2}$)& $ \displaystyle M_2 = \sum_{uv \in E(G)} deg_u deg_v$ & second\_Zagreb()\\
					3 & Hyper Zagreb  ($HM$) & $ \displaystyle HM = \sum_{uv \in E(G)} (deg_u + deg_v)^2$ & hy\_Zagreb()  \\
					4 & Third Zagreb  ($M_{3}$) &  $ \displaystyle M_3 = \sum_{uv \in E(G)} (deg_u - 1)(deg_v - 1)$ & third\_Zagreb()\\ 
					5 & Reduced Zagreb  ($RM$)& $ \displaystyle RM = \sum_{uv \in E(G)} (deg_u - 1)(deg_v - 1)$ & redu\_Zagreb() \\ 
					6&Second modegified Zagreb  ($SM$)& $\displaystyle SM = \sum_{uv \in E(G)} \dfrac{1}{deg_udeg_v}$ &second\_modi\_Zagreb() \\
					7& Randi\'c  ($R$)& $ \displaystyle R = \sum_{uv \in E(G)} \left[ \frac{1}{\sqrt{deg_u deg_v}} \right]$ & Randi\'c() \\
					8& Reciprocal Randi\'c  ($RR$)& $ \displaystyle RR = \sum_{uv \in E(G)} \sqrt{deg_u deg_v}$ &reci\_Randi\'c()\\ 
					9& Reduced reciprocal Randi\'c  ($RRR$)& $\displaystyle RRR = \sum_{uv \in E(G)} \sqrt{(deg_u - 1)(deg_v - 1)}$ &redu\_reci\_Randi\'c() \\
					10 & General Randi\'c  ($GR$)& $\displaystyle GR = \sum_{uv \in E(G)}(deg_udeg_v)^{-1}$ & gen\_Randi\'c() \\
					11 & Atom bond connectivity  ($ABC$)& $\displaystyle ABC = \sum_{uv \in E(G)}  \sqrt{\frac{deg_u+deg_v-2}{deg_udeg_v}}$& atm\_bnd\_connect() \\
					12 & Geometric Arithematic  ($GA$)& $ \displaystyle GA=\sum_{uv \in E(G)}\frac{2\sqrt{deg_udeg_v}}{deg_u+deg_v}$ & geom\_arith() \\
					13 & Harmonic  ($H$)& $\displaystyle H = \sum_{uv \in E(G)} \frac{2}{deg_u + deg_v}$ & harm() \\ 
					14& Sum-connectivity  ($SC$)& $\displaystyle SC = \sum_{uv \in E(G)} \frac{1}{\sqrt{deg_u + deg_v}}$ & sum\_connect() \\
					15& Inverse sum  ($IS$)& $\displaystyle IS = \sum_{uv \in E(G)} \dfrac{deg_udeg_v}{deg_u+deg_v}$ &inv\_sum() \\
					16& Alberston  ($Al$)& $\displaystyle Al = \sum_{uv \in E(G)} \lvert deg_u - deg_v \rvert$ &alberst() \\ 
					17 & Symmetric division  ($SD$)& $\displaystyle SD = \sum_{uv \in E(G)} \dfrac{(deg_u)^{2}+(deg_v)^{2}}{deg_udeg_v}$ &symm\_div() \\
					18& Forgotten  ($F$) & $\displaystyle F = \sum_{uv \in E(G)} deg_u^2 + deg_v^2$ & forgot() \\  	
					19& Sombor  ($So$) & $\displaystyle So = \sum_{uv \in E(G)} \sqrt{deg_u^2 + deg_v^2}$ &somb() \\
					20& Bi-Zagreb  ($BM$) & $\displaystyle BM = \sum_{uv \in E(G)} deg_u + deg_v + deg_udeg_v$& biZagreb() \\
					21 & Tri-Zagreb  ($TM$) & $\displaystyle TM = \sum_{uv \in E(G)} deg_u^2 + deg_v^2 + deg_udeg_v$ & TriZagreb() \\
					22 & Geometric Harmonic  ($GH$)& $ \displaystyle GH=\sum_{uv \in E(G)} (deg_u^2 + deg_v^2 + deg_udeg_v)$ & GeomHarm()\\
					23 & Geometric Bi-Zagreb  ($GBM$) & $\displaystyle GBM=\sum_{uv \in E(G)} \frac{\sqrt{deg_udeg_v}}{deg_u+deg_v+deg_udeg_v}$ & GeomBiZagreb()\\
					24 & Geometric Tri-Zagreb  ($GTM$)&$\displaystyle GTM=\sum_{uv \in E(G)} \frac{\sqrt{deg_udeg_v}}{deg_u^2+deg_v^2+deg_udeg_v}$&  GeomTriZagreb()\\
					25 & Harmonic Geometric  ($HG$) & $\displaystyle HG=\sum_{uv \in E(G)} \frac{2}{\sqrt{deg_udeg_v}(deg_udeg_v)}$ & HarmGeom()\\
					\hline
				\end{tabular}
			}
		\end{table}
		
		\begin{table}[H]
			\centering
			\caption{Neighborhood Degree-based Topological indices, its formulas and its functions}\label{TIformula2}
			\resizebox{6in}{4.6in}{
				\begin{tabular}{|l|l|l|l|}
					\hline 	
					26 & Neighborhood First Zagreb  ($NM_{1}$)& $ \displaystyle NM_1 = \sum_{uv \in E(G)} (w_u + w_v)$ & first\_Zagreb()\\
					27 & Neighborhood second Zagreb  ($NM_{2}$)& $ \displaystyle NM_2 = \sum_{uv \in E(G)} w_u w_v$ & second\_Zagreb()\\
					28 & Neighborhood Hyper Zagreb  ($NHM$) & $ \displaystyle NHM = \sum_{uv \in E(G)} (w_u + w_v)^2$ & hy\_Zagreb()  \\
					29 & Neighborhood Third Zagreb  ($NM_{3}$) &  $ \displaystyle NM_3 = \sum_{uv \in E(G)} (w_u - 1)(w_v - 1)$ & third\_Zagreb()\\ 
					30 & Neighborhood Reduced Zagreb  ($NRM$)& $ \displaystyle NRM = \sum_{uv \in E(G)} (w_u - 1)(w_v - 1)$ & redu\_Zagreb() \\ 
					31 & Neighborhood second modified Zagreb  ($NSM$)& $\displaystyle NSM = \sum_{uv \in E(G)} \dfrac{1}{w_uw_v}$ &second\_modi\_Zagreb() \\
					32 & Neighborhood Randi\'c  ($NR$)& $ \displaystyle NR = \sum_{uv \in E(G)} \left[ \frac{1}{\sqrt{w_u w_v}} \right]$ & Randi\'c() \\
					33 & Neighborhood Reciprocal Randi\'c  ($NRR$)& $ \displaystyle NRR = \sum_{uv \in E(G)} \sqrt{w_u w_v}$ &reci\_Randi\'c()\\ 
					34 & Neighborhood Reduced reciprocal Randi\'c  ($NRRR$)& $\displaystyle NRRR = \sum_{uv \in E(G)} \sqrt{(w_u - 1)(w_v - 1)}$ &redu\_reci\_Randi\'c() \\
					35  & Neighborhood General Randi\'c  ($NGR$)& $\displaystyle NGR = \sum_{uv \in E(G)}(w_uw_v)^{-1}$ & gen\_Randi\'c() \\
					36 & Neighborhood Atom bond connectivity  ($NABC$)& $\displaystyle NABC = \sum_{uv \in E(G)}  \sqrt{\frac{w_u+w_v-2}{w_uw_v}}$& atm\_bnd\_connect() \\
					37 & Neighborhood Geometric Arithematic  ($NGA$)& $ \displaystyle NGA=\sum_{uv \in E(G)}\frac{2\sqrt{w_uw_v}}{w_u+w_v}$ & geom\_arith() \\
					38 & Neighborhood Harmonic  ($NH$)& $\displaystyle NH = \sum_{uv \in E(G)} \frac{2}{w_u + w_v}$ & harm() \\ 
					39 & Neighborhood sum-connectivity  ($NSC$)& $\displaystyle NSC = \sum_{uv \in E(G)} \frac{1}{\sqrt{w_u + w_v}}$ & sum\_connect() \\
					40 & Neighborhood Inverse sum  ($NIS$)& $\displaystyle NIS = \sum_{uv \in E(G)} \dfrac{w_uw_v}{w_u+w_v}$ &inv\_sum() \\
					41& Neighborhood Alberston  ($NAl$)& $\displaystyle NAl = \sum_{uv \in E(G)} \lvert w_u - w_v \rvert$ &alberst() \\ 
					42 & Neighborhood symmetric division  ($NSD$)& $\displaystyle NSD = \sum_{uv \in E(G)} \dfrac{(w_u)^{2}+(w_v)^{2}}{w_uw_v}$ &symm\_div() \\
					43& Neighborhood Forgotten  ($NF$) & $\displaystyle NF = \sum_{uv \in E(G)} w_u^2 + w_v^2$ & forgot() \\  	
					44& Neighborhood sombor  ($NSo$) & $\displaystyle NSo = \sum_{uv \in E(G)} \sqrt{w_u^2 + w_v^2}$ &somb() \\
					45& Neighborhood Bi-Zagreb  ($NBM$) & $\displaystyle NBM = \sum_{uv \in E(G)} w_u + w_v + w_uw_v$ & biZagreb() \\
					46 & Neighborhood Tri-Zagreb  ($NTM$) & $\displaystyle NTM = \sum_{uv \in E(G)} w_u^2 + w_v^2 + w_uw_v$ & TriZagreb() \\
					47 & Neighborhood Geometric Harmonic  ($NGH$)& $ \displaystyle NGH=\sum_{uv \in E(G)} (w_u^2 + w_v^2 + w_uw_v)$ & GeomHarm()\\
					48 & Neighborhood Geometric Bi-Zagreb  ($NGBM$) & $\displaystyle NGBM=\sum_{uv \in E(G)} \frac{\sqrt{w_uw_v}}{w_u+w_v+w_uw_v}$ & GeomBiZagreb()\\
					49 & Neighborhood Geometric Tri-Zagreb  ($NGTM$)&$\displaystyle NGTM=\sum_{uv \in E(G)} \frac{\sqrt{w_uw_v}}{w_u^2+w_v^2+w_uw_v}$&  GeomTriZagreb()\\
					50 & Neighborhood Harmonic Geometric  ($NHG$) & $\displaystyle NHG=\sum_{uv \in E(G)} \frac{2}{\sqrt{w_uw_v}(w_uw_v)}$ & HarmGeom()\\
					\hline
				\end{tabular}
			}
		\end{table}
	\end{enumerate}
	
	Based on the above features of PyTIE, there are four packages and four modules. PyTIE’s packages include class objects, functions, variables, lists, data types, if-else conditions, for loops, and summation. Two types of libraries, namely NumPy, Math and SymPy, are utilized to develop PyTIE’s modules. In the future, we plan to incorporate additional libraries to enhance the extensions of this work.
	\subsection{PyTIE Structures}
	PyTIE has been developed to generate expressions and numerical values for topological indices in single-parameter structures. This discovery holds significant value for chemists and mathematical chemists alike. It implies that the time complexity of PyTIE is $O(1)$, aligning with the efficient computational principles outlined in the theory of time complexity within the molecular descriptors field.
	Comprising four distinct software packages, PyTIE encompasses $\text{PyTIE\_D}$, $\text{PyTIE\_DS}$, $\text{PyTIE\_SMS\_DE}$, and $\text{PyTIE\_SMS\_DSE}$. Each package houses a singular module dedicated to various types of molecular descriptors computation. The execution of these modules involves calling four functions: topoexpressd, topoexpressds, smstopoexpressd, and smstopoexpressds. These functions portray molecular graph representations in terms of minimum degree, maximum degree, and edge partitions, alongside methods for computing diverse graph invariants and topological indices expressions and numerical values.
	To initiate this PyTIE packages in PiP in future, one must employ specific Python commands using PiP:
	\begin{itemize}
		\item For calculating degree-based topological indices of a single-parameter structure: \\
		$>>$import $\text{PyTIE\_D}$ as pd \\
		$>>$from $\text{PyTIE\_D}$ import topological\_indices\_Degree\\
		\hspace*{3cm} or \\
		$>>$import $\text{PyTIE\_D}$ as pd \\
		$>>$from $\text{PyTIE\_D}$ import *
		\item	For calculating degree-sum-based topological indices of a single-parameter structure: \\
		$>>$import $\text{PyTIE\_DS}$ as pds \\
		$>>$from $\text{PyTIE\_DS}$ import topological\_indices\_DegreeSum \\
		\hspace*{3cm} or \\
		$>>$import $\text{PyTIE\_DS}$ as pds \\
		$>>$from $\text{PyTIE\_DS}$ import *
		\item	For calculating degree and entropy-based topological indices of a single molecular graph: \\
		$>>$import $\text{PyTIE\_SMS\_DE}$ as pde \\
		$>>$from $\text{PyTIE\_SMS\_DE}$ import sms\_topological\_indices\_Degree \\
		\hspace*{3cm} or \\
		$>>$import $\text{PyTIE\_SMS\_DE}$ as pde \\
		$>>$from $\text{PyTIE\_SMS\_DE}$ import *
		\item	For calculating degree sum and entropy-based topological indices of a single molecular graph: \\
		$>>$import $\text{PyTIE\_SMS\_DSE}$ as pdse \\
		$>>$from $\text{PyTIE\_SMS\_DSE}$ import sms\_topological\_indices\_DegreeSum \\
		\hspace*{3cm} or \\
		$>>$import $\text{PyTIE\_SMS\_DSE}$ as pdse \\
		$>>$from $\text{PyTIE\_SMS\_DSE}$ import *
	\end{itemize}
	Upon issuing these commands, PyTIE’s packages and modules are automatically imported and allowing researchers to provide input and obtain output for topological indices expressions and its numerical values. For structures involving multiple parameters, the scope of the work will be expanded.
	\subsection{Utilizing PyTIE}
	The utilization of PyTIE is made available through a subscription acquired via its official homepage at PyTIE homepage. Throughout its development, PyTIE underwent testing on Windows OS with Python 3.11.1. It is pertinent to highlight that, lacking compiled code, PyTIE exhibits operating system independence, rendering it applicable on any computer equipped with a Python interpreter. Researchers can refer to the following steps for employing PyTIE after publication on the webpage:
	\begin{itemize}
		\item Access the local file system by opening the HTML file located at PyTIE homepage.
		
		\item Navigate to the PyTIE link within the opened webpage.
		
		\item Complete the registration process and log in to the homepage of PyTIE.
		
		\item Select desired degree-based expressions from the following options:  \\  
		\text{PyTIE\_D}, \text{PyTIE\_DS}, \text{PyTIE\_SMS\_DE}, \text{PyTIE\_SMS\_DSE}
		
		\item Provide a molecular graph as input and initiate the calculation by clicking on the "Get Output" button. It's algorithm is given below:
		\subitem {\bf Algorithm:}
		\subitem {\bf Input:} get minimum degree
		\subitem {\bf Input:} get maximum degree
		\subitem {\bf Input:} specify quantity of edge partitions
		\subitem print \{ edge partitions
		\subitem        \}                       
		\subitem {\bf Input:} enter the number of numerical values needed
		\subitem print \{ topological indices expressions
		\subitem        \}
		\subitem print \{ numerical values of topological indices expressions
		\subitem        \}
		\item Retrieve the resulting topological indices expressions and its numerical values.
	\end{itemize}
	PyTIE depends only on the NumPy, Math and SymPy packages, which may be preinstalled with Python \cite{pytie26}.
	\subsection{Topological indices, neighborhood topological indices and Shannon entropy}
	\subsubsection{Topological indices:}
	Topological indices serve as numeric descriptors in the realm of mathematical chemistry, offering a means to characterize the topology of molecular structures. The origin of topological indices traces back to the mid-20th century when mathematicians and chemists initiated exploration into methods for quantifying molecular structure \cite{pytie27}. In 1971, Randić introduced the molecular connectivity index, among the earliest topological indices. Subsequently, numerous indices have emerged, each emphasizing distinct aspects of molecular structure. Research on topological indices has broadened to encompass a range of applications, such as quantitative structure activity relationships, drug design, and material science. Ongoing efforts by scientists involve refining existing indices and creating novel ones to boost their predictive capabilities and applicability across diverse domains. A comprehensive examination of topological indices would unveil an intricate landscape of mathematical formulations and their respective applications. Researchers frequently delve into the connections between topological indices and molecular properties, with the aim of revealing meaningful correlations that contribute to a deeper comprehension of molecular behavior \cite{pytie28,pytie29}.
	\subsubsection{Neighborhood topological descriptors:}
	Within mathematical chemistry, neighbourhood topological indices are a set of numerical descriptors that are specifically designed to capture the structural characteristics of molecules in their local surroundings. These indices provide important details on the surroundings and the connectivity of a molecule in its immediate environment. Investigating these indices further allowed researchers to address the need for more accurate descriptors that focused on the local properties of molecular structures. To increase the precision of molecular structure analysis, this research acquired impetus in the latter half of the 20th century. The goal is to customize these indices to offer intricate ideas into the local interactions and characteristics of molecules \cite{pytie30,pytie31,pytie32}. A survey of the literature on neighbourhood topological indices reveals a vibrant field in which scholars continuously improve upon and introduce new indices. Often, the main goal is to understand how local structural characteristics affect the behaviour and properties of molecules. PyTIE's contains the computation of topological indices based on degree and neighbourhood from the Tables \ref{TIformula1} and \ref{TIformula2}.
	\subsubsection{Shannon entropy:}
	Claude Shannon's entropy, developed in the late 1940s, marks a pivotal contribution to information theory. In 1948, Shannon introduced the concept of entropy in his groundbreaking paper "A Mathematical Theory of Communication" \cite{pytie52}. 
	Shannon's entropy quantifies the uncertainty or information content associated with a message or data source. Over time, this concept has been applied across diverse domains, such as cryptography, data analysis, and statistical mechanics \cite{pytie43}. Entropy has broad applicability across multiple domains, including materials science, optoelectronics, nanotechnology, energy storage, sensing technologies, and electronics \cite{pytie58,pytie60}. It became a fundamental concept in information theory, influencing diverse disciplines such as computer science, telecommunications, and mathematics. Also PyTIE's contains the computation of the following degree based and neighborhood based entropy. These are obtained the formulas from Table \ref{TIformula1} by using Eqn. 1. 
	\subsection{Examples of PyTIE}
\begin{figure}[H]
		\centering
		\subfloat{\includegraphics[width=12cm]{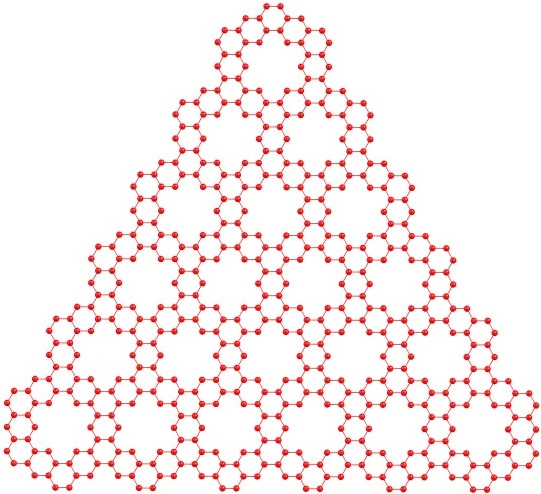}\label{Stagen}}
		\quad \quad
		\caption{Molecular graph of tessellations of kekulenes (G)}\label{G}
	\end{figure}

\begin{figure}[H]
		\centering
		\subfloat[]{\includegraphics[width=4cm]{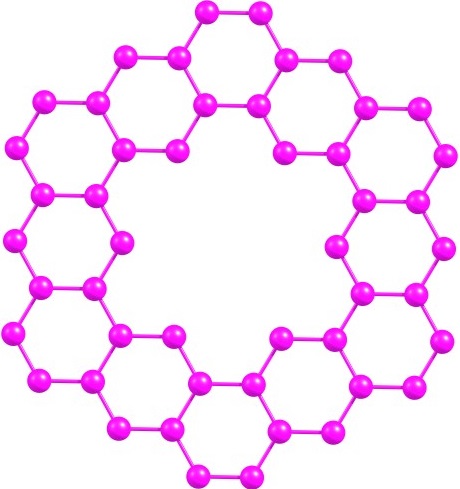}\label{Stage1}}
		\quad \quad
		\subfloat[]{\includegraphics[width=6cm]{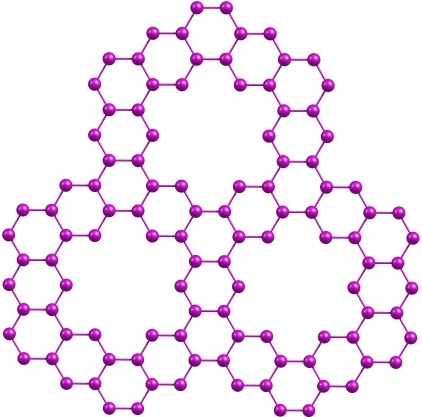}\label{Stage2}}
		\quad \quad
		\caption{Molecular graph of tessellations of kekulenes a) Stage-1 of G b) Stage-2 of G}\label{stg12}
	\end{figure}

\begin{figure}[H]
		\centering
		\subfloat[]{\includegraphics[width=10cm]{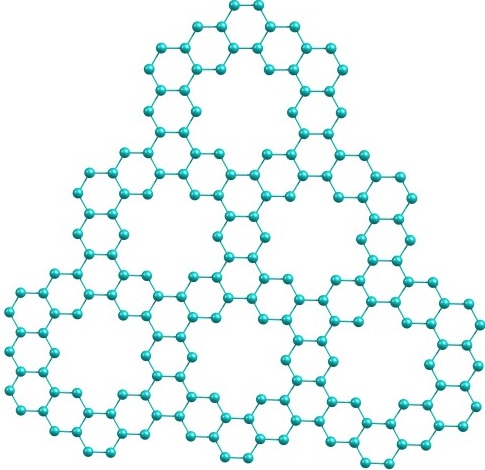}\label{Stage3}}
		\quad \quad
		\subfloat[]{\includegraphics[width=12cm]{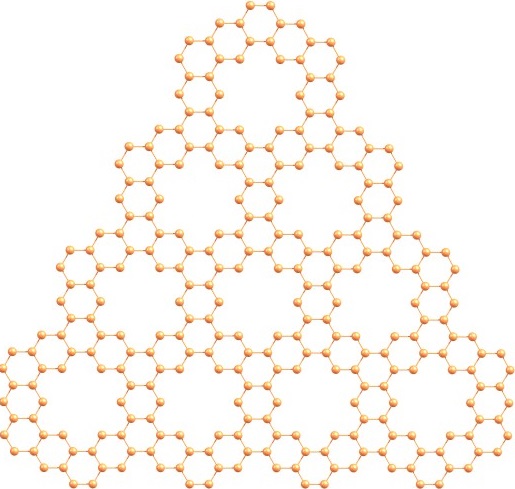}\label{Stage4}}
		\quad \quad
		\caption{Molecular graph of tessellations of kekulenes a) Stage-3 of G b) Stage-4 of G }\label{stg34}
	\end{figure}

\begin{figure}[H]
		\centering
		\subfloat{\includegraphics[width=9cm]{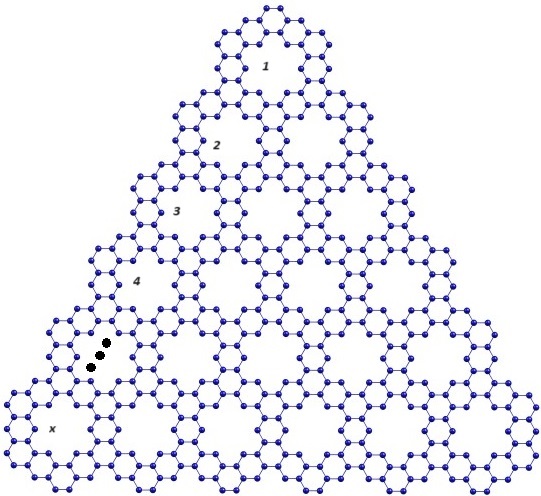}\label{stagen}}
		\quad \quad
		\caption{Molecular graph of tessellations of kekulenes (Stage-n of G)}\label{stgn}
	\end{figure}

	In this investigation, the outcomes pertaining to topological indices of tessellations involving Triangular kekulene (\(G\)) (See Figure \ref{G}) is juxtaposed with the computational results obtained through our proprietary software. The studies conducted on \(G\) can be referenced in the work by Govardhan et al \cite{pytie44}. Additional details about the Triangular kekulene structure in question can be acquired from the aforementioned article \cite{pytie44}.
	We demonstrate the high computational efficiency of PyTIE software through empirical examples. Comparative analysis reveals that manual computations for the stages of the graph $G$ (See Figures \ref{stg12}-\ref{stgn}), performed within a linear time framework, necessitate more time than the corresponding computations executed by PyTIE for the same graph $G$.
	Execute the PyTIE modules by providing input parameters, including the lowest and highest degree, the total number of edge pairs (edge partitions), the overall number of partitions, and the required number of numerical values (N.V) for each topological index. Upon receiving these inputs, the PyTIE modules will generate the output for topological indices expressions and its numerical values and corresponding numerical values, achieving constant time complexity.
	\subsubsection{Python module of Degree-based topological indices of $G$}
	{\small \small  \bf $\text{PyTIE\_D}$'s input of $G$:}
	\begin{table}[H]
		\centering
		\resizebox{3.5in}{1.6in}{
			\begin{tabular}{|l|}
				\hline
				\textcolor{red}{\bf Let's Compute Degree-based Topological Descriptors of $G$ } \\
				\textcolor{blue}{\bf Provide your data to calculate Degree-based Topological Descriptors of $G$:} \\
				$>>$Specify the \textcolor{magenta}{\bf minimum degree} of a graph \textcolor{green}{\bf G$_{1}$}  : \textcolor{red}{\bf 2} \hspace*{5cm} \\
				$>>$Specify the \textcolor{magenta}{\bf maximum degree} of a graph \textcolor{green}{\bf $G$}  : \textcolor{red}{\bf 3} \\
				$>>$Specify the \textcolor{green}{\bf partition} value of \textcolor{blue}{1} dimension of \textcolor{blue}{(2, 2)} of \textcolor{green}{$G$}  : \textcolor{red}{3} \\
				$>>$Specify the \textcolor{green}{\bf partition} value of \textcolor{blue}{2} dimension of \textcolor{blue}{(2, 2)} of \textcolor{green}{$G$}  :  \textcolor{red}{6}\\
				$>>$Specify the \textcolor{green}{\bf partition} value of \textcolor{blue}{3} dimension of \textcolor{blue}{(2, 2)} of \textcolor{green}{$G$}  : \textcolor{red}{9} \\
				$>>$Specify the \textcolor{green}{\bf partition} value of \textcolor{blue}{4} dimension of \textcolor{blue}{(2, 2)} of \textcolor{green}{$G$}  : \textcolor{red}{12}\\
				$>>$Specify the \textcolor{green}{\bf partition} value of \textcolor{blue}{5} dimension of \textcolor{blue}{(2, 2)} of \textcolor{green}{$G$}  : \textcolor{red}{15}\\
				$>>$Specify the \textcolor{green}{\bf partition} value of \textcolor{blue}{6} dimension of \textcolor{blue}{(2, 2)} of \textcolor{green}{$G$}  : \textcolor{red}{18} \\
				$>>$Specify the \textcolor{green}{\bf partition} value of \textcolor{blue}{1} dimension of \textcolor{blue}{(2, 3)} of \textcolor{green}{$G$}  : \textcolor{red}{36}\\
				$>>$Specify the \textcolor{green}{\bf partition} value of \textcolor{blue}{2} dimension of \textcolor{blue}{(2, 3)} of \textcolor{green}{$G$}  : \textcolor{red}{78}\\
				$>>$Specify the \textcolor{green}{\bf partition} value of \textcolor{blue}{3} dimension of \textcolor{blue}{(2, 3)} of \textcolor{green}{$G$}  : \textcolor{red}{132}\\
				$>>$Specify the \textcolor{green}{\bf partition} value of \textcolor{blue}{4} dimension of \textcolor{blue}{(2, 3)} of \textcolor{green}{$G$}  : \textcolor{red}{198}\\
				$>>$Specify the \textcolor{green}{\bf partition} value of \textcolor{blue}{5} dimension of \textcolor{blue}{(2, 3)} of \textcolor{green}{$G$}  : \textcolor{red}{276}\\
				$>>$Specify the \textcolor{green}{\bf partition} value of \textcolor{blue}{6} dimension of \textcolor{blue}{(2, 3)} of \textcolor{green}{$G$}  : \textcolor{red}{366}\\
				$>>$Specify the \textcolor{green}{\bf partition} value of \textcolor{blue}{1} dimension of \textcolor{blue}{(3, 3)} of \textcolor{green}{$G$}  : \textcolor{red}{15}\\
				$>>$Specify the \textcolor{green}{\bf partition} value of \textcolor{blue}{2} dimension of \textcolor{blue}{(3, 3)} of \textcolor{green}{$G$}  : \textcolor{red}{48}\\
				$>>$Specify the \textcolor{green}{\bf partition} value of \textcolor{blue}{3} dimension of \textcolor{blue}{(3, 3)} of \textcolor{green}{$G$}  : \textcolor{red}{93}\\
				$>>$Specify the \textcolor{green}{\bf partition} value of \textcolor{blue}{4} dimension of \textcolor{blue}{(3, 3)} of \textcolor{green}{$G$}  : \textcolor{red}{150}\\
				$>>$Specify the \textcolor{green}{\bf partition} value of \textcolor{blue}{5} dimension of \textcolor{blue}{(3, 3)} of \textcolor{green}{$G$}  : \textcolor{red}{219}\\
				$>>$Specify the \textcolor{green}{\bf partition} value of \textcolor{blue}{6} dimension of \textcolor{blue}{(3, 3)} of \textcolor{green}{$G$}  : \textcolor{red}{300}\\
				$>>$Specify the total number of \textcolor{green}{\bf partitions}  : \textcolor{red}{3} \\
				\hline
			\end{tabular}
		}
	\end{table}
	
	{ \small \small \bf $\text{PyTIE\_D}$'s output of $G$:}
	\begin{table}[H]
		\centering
		
		\begin{tabular}{|l|}
			\hline
			\subfloat{\includegraphics[width=17cm]{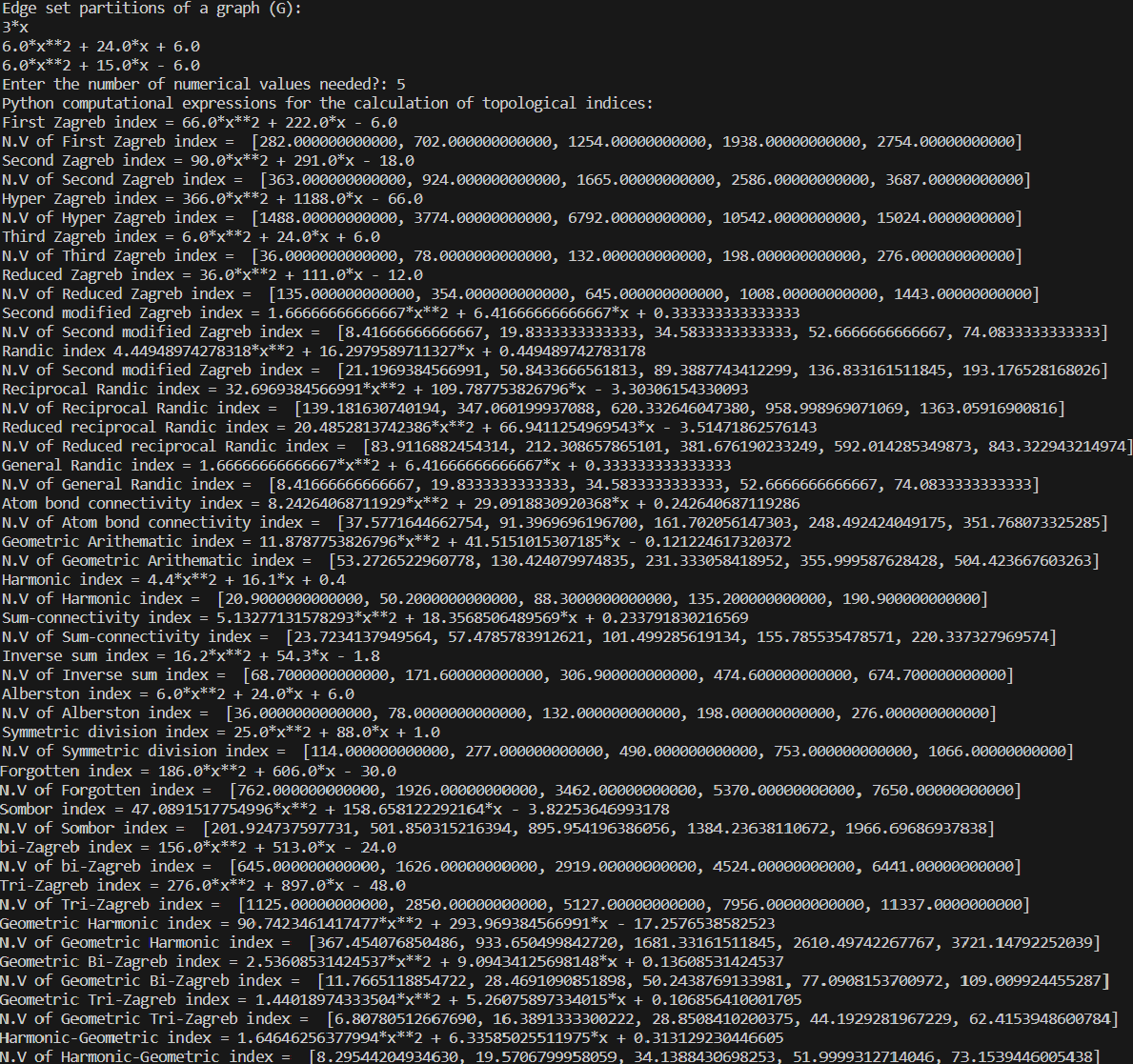}\label{BA}} \\
			\hline
		\end{tabular}
	\end{table}
	The structured outputs, as generated by the aforementioned examples of PyTIE modules, encompass a substantial repository of topological data, which holds potential for application in machine learning to scrutinize the spectroscopies and stabilities of diverse molecular structures. Graph spectra and additional measurements, such as resonance energy measures per bond, can be computed based on the combinatorial counts of molecular structures utilizing PyTIE software. The derived data can be leveraged to estimate both thermodynamic and kinetic stabilities across a range of molecular structures; this possibility is currently under consideration. Furthermore, the outcomes furnished by PyTIE modules can be employed to formulate meaningful quantitative predictions regarding the molecular properties of compounds. In the upcoming section, we have given example one of the software package such as PyTIE\_D. Like wise, the other software packages also can compute the results.
	\section{Robust Calculations of Enthalpies and Relative Stabilities using Bond (Edge) Distributions of Triangular and Rhomboidal Kekulenes}
	
	This section outlines the use of additive degree-based edge distributions to determine the enthalpies and relative stabilities of triangular kekulene. While these distributions provide greater discriminatory power compared to traditional degree based bond distributions, they are distinct from automorphic bond distributions. We utilize degree sum based bond distributions to effectively capture higher-dimensional characteristics of triangular kekulene by leveraging the additive behavior of bond enthalpies. Consequently, edge partition classes present an accurate and efficient approach for evaluating thermodynamic properties, including formation enthalpies and Gibbs free energies. In the context of molecular graph theory, the carbon framework of the molecule, composed of \ce{C}-\ce{C} bonds, is typically employed. In the present analysis, however, both \ce{C}-\ce{C} and \ce{C}-\ce{H} bonds in the kekulene tessellations are considered, resulting in the subsequent general formulations for the total formation enthalpies of these tessellated structures.\\
	
	$\Delta \ce{H}_{f}^{298K}(G)=\displaystyle\sum_{i=1}^{N_{e}}|\ce{C_{i}}|\Delta H_{fi}^{298K}(\ce{C-C)_{i}}+\displaystyle\sum_{i=1}^{N_{h}}|\ce{C_{j}}|\Delta H_{fj}^{298K}(\ce{C-H)_{j}}$
	\begingroup
	\allowdisplaybreaks
	\begin{align*}
		\Delta H_{f}^{298K}(G_{T}(n))&=(3n+3)\Delta H_{f1}^{298K}(\ce{C_{5}-C_{5}})+(6n+6)\Delta H_{f2}^{298K}(\ce{C_{5}-C_{7}})& \\ & \ \ +(6n+6)\Delta H_{f3}^{298K}(\ce{C_{6}-C_{7}})+(6n^{2}+12n-6)\Delta H_{f4}^{298K}(\ce{C_{6}-C_{8}})& \\ & \ \ +(6n+6)\Delta H_{f5}^{298K}(\ce{C_{7}-C_{8}})+(6n^{2}+9n-9)\Delta H_{f6}^{298K}(\ce{C_{8}-C_{8}})& \\ & \ \ +(12n^{2}+48n+12)\Delta H_{fC-H}^{298K}(\ce{C-H)}&\\
		\Delta H_{f}^{298K}(G_{T}(4))&=(15)\Delta H_{f1}^{298K}(\ce{C_{5}-C_{5}})+(30)\Delta H_{f2}^{298K}(\ce{C_{5}-C_{7}})& \\ & \ \ +(30)\Delta H_{f3}^{298K}(\ce{C_{6}-C_{7}})+(138)\Delta H_{f4}^{298K}(\ce{C_{6}-C_{8}})& \\ & \ \ +(30)\Delta H_{f5}^{298K}(\ce{C_{7}-C_{8}})+(123)\Delta H_{f6}^{298K}(\ce{C_{8}-C_{8}}
		)& \\ & \ \ +(396)\Delta H_{fC-H}^{298K}(\ce{C-H})&
	\end{align*}
	Similar methods can be used to obtain expressions for the enthalpies of formation for various kekulenes. The free energies of these structures can also be calculated similarly.
	\section{Conclusion}
	We have successfully devised PyTIE (Python Topological Indices Expressions), a suite of Python packages designed for the computation of topological indices expressions and its numerical values. The optimal instances have been presented to illustrate the procedural aspects of PyTIE's computational software. These illustrations elucidate the equivalence between manual computations and PyTIE's computational results. The optimization of the computation of topological indices has been achieved through the creation of an open-source Python tool named PyTIE. This tool integrates advanced analysis methods, including algorithmic differentiation and constraints derived from property prediction analyses. Algorithmic differentiation is utilized to derive expressions and numerical values for diverse topological indices, resulting in a reduction in both gradient estimation costs and time, especially in situations involving a high number of parameters. The edge partitions based on degree sum are useful in estimating the heat of formation and enthalpy of formation along with Gaussian G-2 theory. These improvements collectively increase the viability of performing top-down identifications of different chemical properties and data. Potential future enhancements for PyTIE could include improving its capacity to handle larger datasets and complicated molecular structures, as well as integrating machine learning for predictive modelling based on topological indices. These developments are intended to speed up predictions of material properties and make it easier to find new materials for a variety of scientific uses. Additionally, PyTIE's software exhibits proficiency in addressing intricate problems involving multiple parameter structures of chemical graphs. PyTIE's software packages enable researchers to rapidly predict molecular properties of molecule structures through the utilization of topological descriptors. We welcome contributions and implementation requests through the https://github.com/sahayavijay/PyTIE-0.0.1.git.
	
	\subsubsection*{Data and Software Availability}
	The PyTIE packages available open source, free of charge, on https://github.com/sahayavijay/PyTIE-0.0.1.git.
	
	\subsubsection*{Acknowledgement}
	The authors express their sincere gratitude to Vellore Institute of Technology, Vellore, for providing financial support.

\end{document}